\documentstyle[multicol,aps,psfig]{revtex}
\begin{document}
\draft
\preprint{LBNL-48325}
 
\title{Parton Energy Loss with Detailed Balance}

\author{Enke Wang$^a$ and Xin-Nian Wang$^{a,b}$}

\address{$^a$Institute of Particle Physics, Huazhong Normal University,
         Wuhan 430079, China}
\address{$^b$Nuclear Science Division, Lawrence Berkeley Laboratory, 
         Berkeley, California 94720}

\date{June 15, 2001}

\maketitle

\begin{abstract}
Stimulated gluon emission and thermal absorption in addition to 
induced radiation are considered for an energetic parton propagating 
inside a quark-gluon plasma. In the presence of thermal gluons, 
stimulated emission reduces while absorption increases the parton's 
energy. The net effect is a reduction of the parton energy loss. 
Though decreasing asymptotically as $T/E$ with the 
parton energy, the relative reduction is found to be important for 
intermediate energies. The modified energy dependence of the 
energy loss will affect the shape of suppression of moderately 
high $p_T$ hadrons due to jet quenching in high-energy heavy-ion collisions.

\end{abstract}


\begin{multicols}{2}

{\it Introduction.}--- Gluon radiation induced by multiple scattering
for an energetic parton propagating in a dense medium leads to medium
induced parton energy loss or jet quenching. Such a phenomenon, if
occurring in high-energy heavy-ion collisions, will suppress large
transverse momentum hadrons as compared to $pp$ collisions at the
same energy \cite{WG92}.  Since medium-induced energy loss is 
proportional to the gluon density \cite{GW94}, the experimental studies of 
high $p_T$ pion suppression \cite{QM01D,QM01W} in high-energy heavy-ion 
collisions will provide a direct measurement of the initial conditions
of the produced dense matter.

Recent theoretical studies \cite{BDPMS,Zakharov,GLV,GuoW,Wied} of parton
energy loss have concentrated on gluon radiation induced by multiple
scattering in a medium. Such a radiative energy loss demonstrates many
interesting properties due to the non-Abelian Landau-Pomeranchuk-Migdal (LPM)
interference effect \cite{LPM}. Since gluons are bosons, there should also
be stimulated gluon emission and absorption by the propagating parton 
because of the presence of thermal gluons in the hot medium.  Such detailed
balance is crucial for parton thermalization and should also be important
for calculating the energy loss of an energetic parton in a hot medium.

In this Letter, we report a first study of the effect of stimulated emission
and thermal absorption on the energy loss of a propagating parton in a hot QCD
medium. We consider both the final-state radiation associated with the hard
processes that have produced the original hard parton and the radiation
induced by final-state multiple scattering in the medium. Naively, the 
energy scale associated with stimulated emission and thermal absorption 
should both be around the temperature of the medium, $\omega \sim T$.
However, we will show in this Letter that the partial cancellation between 
stimulated emission and thermal absorption results in a net reduction 
of parton energy loss induced by multiple scattering.
The relative reduction decreases with the parton energy $E$ as
$T/E$, as a consequence of the LPM interference.
Though such a reduction is negligible for asymptotically large parton energy,
it is still important for intermediate values of $E$. It also modifies 
the energy dependence of the total energy loss in the small to 
intermediate energy region, which is very relevant to the jet quenching
phenomenon for intermediate large $p_T<10$ GeV/$c$ hadrons.

{\it Final-state absorption.} ---  Jet production in a hard scattering
is always accompanied by final-state radiation. In an axial gauge and
in the leading-log approximation, the radiation amplitude off a quark
can be factorized from the hard scattering and has the form,
 \begin{equation}
 \label{rad0}
   R^{(0)}=2ig T_c
   {{{\bf k}_{\perp}\cdot{\bf \epsilon}_{\perp}}\over {\bf k}^2_{\perp}}\, ,
 \end{equation}
in the limit of $z=|\omega|/E\rightarrow 0$, where $k=(\omega,{\bf k})$ is
the four-momentum of the radiated gluon with polarization
$\epsilon=[0,({\bf k}_{\perp}\cdot{\bf \epsilon}_{\perp})/k^+,
{\bf \epsilon}_{\perp}]$, $T_c$ the color matrix
and $\alpha_{s}=g^2/4\pi$ strong coupling constant. 
We assume that the hot medium is in thermal equilibrium shortly 
after the production of the hard parton. Therefore, taking into
account of both stimulated emission and thermal absorption in
a thermal medium with finite temperature $T$, one has the
probability of gluon radiation with energy $\omega$,
 \begin{eqnarray}
 \label{prob0}
   {{dP^{(0)}}\over d\omega}&&={{\alpha_s C_F}\over 2\pi}
     \int {{dz}\over z} \int
     {{d{\bf k}_{\perp}^2}\over {\bf k}_{\perp}^2}
     \Big[ N_g(zE)\delta(\omega+zE)
 \nonumber\\
     &&\left(1+N_g(zE)\right)\delta(\omega-zE)\theta(1-z)\Big]
     P({\omega\over E})\, ,
 \end{eqnarray}
where, $N_g(|{\bf k}|)=1/[\exp(|{\bf k}|/T)-1]$ is the thermal gluon
distribution and $C_F$ is the Casimir of the quark jet in the fundamental
representation. We have also included the splitting 
function $P_{gq}(z)\equiv P(z)/z=[1+(1-z)^2]/z$ for $q\rightarrow gq$.
The first term is from thermal absorption and the second term
from gluon emission with the Bose-Einstein enhancement factor.
For $E\gg T$, one can neglect the quantum statistical effect for the
leading parton. Note that the 
vacuum part has a logarithmic infrared divergency
while the finite-temperature part has a linear divergency, since
$N_g(|{\bf k}|)\sim T/|{\bf k}|$ as $|{\bf k}|\rightarrow 0$.
These infrared divergences will be canceled by the virtual 
corrections which also contain a zero-temperature and a finite-temperature 
part \cite{WWlong}.
In addition, the virtual corrections are also essential to ensure
unitarity and momentum conservation in the QCD evolution of the
fragmentation functions \cite{GuoW}. However, they do not contribute to
the effective parton energy loss. The remaining collinear divergency
in the above spectrum can be absorbed into a renormalized fragmentation
function that follows the QCD evolution equations.

To define the effective parton energy loss, we only consider gluon 
radiation outside a cone with $|{\bf k}_{\perp}|>\mu$. Here, $\mu$ should
be a nonperturbative scale for parton fragmentation in the vacuum. In
a hot QCD medium, it can be replaced by the gluon's Debye screening
mass. Assuming the scale of the hard scattering as $Q^2=4E^2$, one 
has the kinematic limits of the gluon's transverse momentum,
 \begin{equation}
 \label{boundary}
   \mu^2 \leq {\bf k}_{\perp max}^2 \leq 4|\omega|(E-\omega) \, .
 \end{equation}

Subtracting the gluon radiation spectrum in the vacuum, one then
obtains the energy loss due to final-state absorption and stimulated
emission,
 \begin{eqnarray}
 \label{eloss0}
   \Delta E^{(0)}_{abs} &=& - \int d\omega \,\omega
   \left({{dP^{(0)}}\over d\omega}-
     {{dP^{(0)}}\over d\omega}\Big|_{T=0}\right)
 \nonumber\\
   &=&  {{\alpha_s C_F}\over 2\pi}E
   \int dz   \int {{d{\bf k}_{\perp}^2}\over {\bf k}_{\perp}^2}
 \Big[ P(-z)N_g(zE)
 \nonumber\\
   && -P(z) N_g(zE) \theta(1-z) \Big].
\end{eqnarray}

Even though the stimulated emission cancels part of the contribution 
from absorption, the net medium effect without rescattering is 
still dominated by the final-state thermal absorption, 
resulting in a net energy gain. 
For asymptotically large parton energy, $E\gg T$, one can 
complete the above integration approximately and have,
 \begin{eqnarray}
 \label{elossab0}
   {\Delta E^{(0)}_{abs}\over E}&\approx&
   {{\pi\alpha_s C_F}\over 3}
   {T^2\over E^2}\left[
     \ln{4ET\over \mu^2}+2-\gamma_{\rm E}
     +{{6\zeta^\prime(2)}\over \pi^2}\right],
\end{eqnarray}
where, $\gamma_{\rm E}\approx 0.5772$ and $\zeta^\prime(2)\approx -0.9376$.
Terms that are proportional to $\exp(-E/T)$ or beyond the order of $(T/E)^2$
are neglected.
The quadratic temperature dependence of the leading contribution
is a direct consequence of the partial cancellation between stimulated 
emission and thermal absorption, each having a leading contribution 
linear in $T$.

{\it Rescattering-induced absorption.} ---
During the propagation of the hard parton after its production, it
will suffer multiple scattering with the medium. The multiple
scattering in turn can also induce gluon radiation which has been
the focus of recent theoretical studies of radiative energy loss
\cite{BDPMS,Zakharov,GLV,GuoW,Wied}. Here we will investigate
the stimulated emission and thermal absorption associated with
multiple scattering in a hot QCD medium.

Assuming a hard parton produced at ${\bf y}_0=(y_0, {\bf y}_{0\perp})$ 
inside the medium with $y_0$ being the longitudinal coordinate, 
we model the interaction between the jet and target partons
by a static color-screened Yukawa potential as
in Gyulassy-Wang (GW) \cite{GW94},
 \begin{eqnarray}
 \label{potential}
   V_n
    &=&2\pi\delta(q^0)v({\bf q_n})e^{-i{\bf q}_n\cdot {\bf y}_n}
       T_{a_n}(j)T_{a_n}(n)\, , \\
 \label{potential1}
  v({\bf q}_n)&=&{{4\pi\alpha_s}\over {{\bf q}^2_n+\mu^2}}\, .
 \end{eqnarray}
Here ${\bf q}_n$ is the momentum transfer from
a target parton $n$ at ${\bf y}_n=(y_n, {\bf y}_{\perp n})$,
 $T_{a_n}(j)$ and $T_{a_n}(n)$ are the color matrices for the jet and
target parton.

We will also follow the framework of opacity expansion developed by 
Gyulassy, L\'evai and Vitev (GLV)\cite{GLV} and Wiedemann \cite{Wied}.
However, we will only consider contributions to the first
order in the opacity expansion. It was shown by GLV that the higher
order corrections contribute little to the radiative energy loss.
The opacity is defined by the mean number of collisions in the 
medium, ${\bar n}\equiv L/\lambda=N\sigma_{el}/A_{\perp}$. 
Here $N$, $L$ and $A_{\perp}$ are the number, thickness and
transverse area of the targets, and $\lambda$ is the average 
mean-free-path for the jet.

The radiation amplitude associated with a single rescattering is \cite{GLV}
 \begin{eqnarray}
 \label{amps}
   R^{(S)}&=&2ig \Bigl( {\bf H}T_a T_c
     +{\bf B}_1 e^{i\omega_0 y_{10}}[T_c, T_a]
 \nonumber\\
   &&+{\bf C}_1 e^{i(\omega_0-\omega_1) y_{10}}[T_c, T_a]\Bigr)
     \cdot \epsilon_{\perp}\, ,
 \end{eqnarray}
where, $y_{10}=y_1-y_0$,
 \begin{eqnarray}
 \label{omega}
   \omega_0&=&{{{\bf k}_{\perp}^2}\over {2\omega}}\, ,
   \quad
   \omega_1={{({\bf k}_{\perp}-{\bf q}_{\perp})^2}\over
      {2\omega}}\, ,
 \\
   {\bf H}&=&{{\bf k}_{\perp}\over {{\bf k}_{\perp}^2}}\, ,
   \quad
   {\bf C}_1={{{\bf k}_{\perp}-{\bf q}_{\perp}}\over
      ({{\bf k}_{\perp}-{\bf q}_{\perp}})^2}\, ,
   \quad
   {\bf B}_1={\bf H}-{\bf C}_1\, .
 \end{eqnarray}
The first, third and second terms correspond to the hard final-state 
radiation, radiation induced by the rescattering and the interference, 
respectively.
 
To the first order in opacity, one should also include the 
interference between the processes of double and no
rescattering. Assuming no color correlation between different targets,
the double rescattering corresponds to the ``contact limit'' 
of double Born scattering with the same target. The radiation amplitude can
be found as
\begin{eqnarray}
 \label{ampd}
   R^{(D)}&=&2ig T_c e^{i\omega_0y_{10}}
   \Bigl(-{{C_F+C_A}\over 2}{\bf H} e^{-i\omega_0y_{10}}
 \nonumber\\
   &&+{C_A\over 2}{\bf B}_1
    +{C_A\over 2}{\bf C}_1 e^{-i\omega_1y_{10}}\Bigr)
     \cdot \epsilon_{\perp}\, .
\end{eqnarray}
Multiplying with the radiation amplitude with no rescattering $R^{(0)}$,
the first term in $R^{(D)}$ cancels exactly the hard radiation contribution
in the single scattering amplitude $R^{(S)}$.

Similarly to the case of final-state absorption, one can also
include stimulated emission and thermal absorption when calculating
the radiation probability at the first order in opacity,
\end{multicols}
\vspace{0.1in}
 \begin{eqnarray}
 \label{prob1}
   {{dP^{(1)}}\over d\omega}=&&{{C_2}\over {8\pi d_A d_R}}
     \int {{dz}\over z}  \int
      {{d^2{\bf k}_{\perp}}\over {(2\pi)^2}}
     \int {{d^2{\bf q}_{\perp}}\over {(2\pi)^2}}
      v^2({\bf q}_{\perp})P({\omega\over E}){N\over A_{\perp}}
      \left\langle Tr\left[|R^{(S)}|^2
     +2 Re\left(R^{(0)\dagger}
     R^{(D)}\right)\right]\right\rangle
 \nonumber\\
     &&\Big[\left(1+N_g(zE)\right)\delta(\omega-zE)\theta(1-z)
     +N_g(zE)\delta(\omega+zE)\Big]
 \nonumber\\
     =&&{{\alpha_s C_2 C_F C_A}\over {d_A \pi}}
     \int {{dz}\over z}  \int
      {{d{\bf k}_{\perp}^2}\over {\bf k}_{\perp}^2}
     \int {{d^2{\bf q}_{\perp}}\over {(2\pi)^2}}
     v^2({\bf q}_{\perp}) P({\omega\over E})
      {{{\bf k}_{\perp}\cdot {\bf q}_{\perp}}\over
      {\left({\bf k}_{\perp} - {\bf q}_{\perp}\right)^2}}
      {N\over A_{\perp}}
      \left\langle Re(1-e^{i\omega_1 y_{10}})\right\rangle
 \nonumber\\
     &&\Big[\left(1+N_g(zE)\right)\delta(\omega-zE)\theta(1-z)
     +N_g(zE)\delta(\omega+zE)\Big] \, ,
 \end{eqnarray}
\vspace{-0.1in}
\begin{multicols}{2}
\noindent where $C_2$ and $C_A$ are Casimirs of the target parton in the
fundamental and adjoint representations in $d_R$ and $d_A$ dimension, 
respectively. The factor $1-\exp(i\omega_1 y_{10})$ reflects the destructive 
interference arising from the non-Abelian LPM effect\cite{LPM}. Averaging
over the longitudinal target profile is defined as $\left\langle
\cdots\right\rangle=\int dy \rho(y)\cdots$. The target distribution
is assumed to be an exponential form $\rho(y)=2 \exp(-2y/L)/L$.

Again, one should in principle include contributions from virtual 
corrections \cite{WWlong} which will cancel the infrared divergences
in the real emission (absorption) processes. However, they do not
contribute to the effective parton energy loss and can be neglected
here. Similarly to the final-state absorption, the contribution from
thermal absorption associated with rescattering is larger than that of
stimulated emission, resulting in a net energy gain. However, the
zero-temperature contribution corresponds to the radiation induced
by rescattering which will lead to an effective energy loss by the
leading parton. We denote this part as $\Delta E^{(1)}_{rad}$ which 
should be the same as obtained by previous studies \cite{GLV}. The
remainder or temperature-dependent part of energy loss induced by 
rescattering at the first order in opacity is then defined as,
\begin{equation}
 \label{eloss1}
   \Delta E^{(1)}_{abs}=-\int d\omega \,\omega 
     \left({{dP^{(1)}}\over d\omega} -{{dP^{(1)}}\over d\omega}
       \Big|_{T=0}\right)
 \end{equation}
which mainly comes from thermal absorption with partial cancellation
by stimulated emission in the medium. According to Eq.~(\ref{prob1}),
\end{multicols}
 \begin{eqnarray}
 \label{elossem1}
   \Delta E_{rad}^{(1)} &&=-
   {{\alpha_s C_F}\over \pi}{L\over \lambda_g}E
   \int dz  \int {d{\bf k}_{\perp}^2\over {\bf k}_{\perp}^2}
   \int d^2{\bf q}_{\perp}
   |{\bar v}({\bf q}_{\perp})|^2
      {{{\bf k}_{\perp}\cdot {\bf q}_{\perp}}\over
      {\left({\bf k}_{\perp} - {\bf q}_{\perp}\right)^2}} P(z)
     \left\langle Re(1-e^{i\omega_1 y_{10}})\right\rangle
      \theta(1-z)\, ,
 \\
 \label{elossab1}
   \Delta E_{abs}^{(1)}&&=
   {{\alpha_s C_F}\over \pi}{L\over \lambda_g}E
   \int dz \int {d{\bf k}_{\perp}^2\over {\bf k}_{\perp}^2}
   \int d^2{\bf q}_{\perp}
    |{\bar v}({\bf q}_{\perp})|^2
      {{{\bf k}_{\perp}\cdot {\bf q}_{\perp}}\over
      {\left({\bf k}_{\perp} - {\bf q}_{\perp}\right)^2}} N_g(zE)
 \nonumber\\
     &&\Bigl[P(-z)\left\langle
     Re(1-e^{i\omega_1 y_{10}})\right\rangle-
     P(z)\left\langle
     Re(1-e^{i\omega_1 y_{10}})\right\rangle\theta(1-z) \Bigr]\, ,
\end{eqnarray}
\begin{multicols}{2}
\noindent where, $\lambda_g=C_F\lambda/C_A$ is the mean-free-path 
of the gluon, and $|{\bar v}({\bf q}_{\perp})|^2$ is the normalized 
distribution of momentum transfer from the scattering centers,
 \begin{eqnarray}
 \label{vbar}
  |{\bar v}({\bf q}_{\perp})|^2 &\equiv& {1\over \sigma_{el}}
  {d^2\sigma_{el}\over d^2{\bf q}_{\perp}}=
  {1\over \pi}{\mu^2_{eff}\over ({\bf q}_{\perp}^2+\mu^2)^2}\, ,
 \\
   {1\over \mu^2_{eff}} &=& {1\over \mu^2}-{1\over
   q_{\perp max}^2+\mu^2}\,\,\, , q_{\perp max}^2\approx 3E\mu.
 \end{eqnarray}

In the limit $q_{\perp max}\rightarrow\infty$, the angular integral 
can be carried out by partial integration \cite{GLV}. These 
contributions to the energy loss become   
 \begin{eqnarray}
 \label{elossem2}
   \Delta E_{rad}^{(1)}\approx &&-
   {{\alpha_s C_F}\over 2\pi}{L\over \lambda_g}E
   \int dz P(z) h(\gamma)\theta(1-z)\, ,
 \\
 \label{elossab2}
   \Delta E_{abs}^{(1)}\approx &&
   {{\alpha_s C_F}\over 2\pi}{L\over \lambda_g}E
   \int dz N_g(zE)  h(\gamma) 
 \nonumber\\
&& \Big[ P(-z) -P(z) \theta(1-z)\Big]\, ,
 \end{eqnarray}
where, $\gamma=\mu^2 L/(4zE)$ and
\begin{equation}
h(\gamma)=\left\{
\begin{array}{ll}
{2\gamma \over \sqrt{1-4\gamma^2}} [{\pi \over 2}-\arcsin(2\gamma)]\, ,
& \gamma<1/2 \\
{2\gamma \over \sqrt{4\gamma^2-1}}\ln[2\gamma+\sqrt{4\gamma^2-1}]\, ,
&  \gamma>1/2\, .
\end{array}
\right.
\end{equation}
One can approximate $h(\gamma)$ with 
$\pi\gamma+(11/4-2\pi)\gamma^2+(5/2)\gamma^3$ for $\gamma<1/2$ and 
$\ln(4\gamma)+0.1/\gamma+0.028/\gamma^2$ for $\gamma>1/2$.
In the limit of $EL\gg 1$ and $E\gg \mu$, One can then get the
approximate asymptotic behavior of the energy loss,
 \begin{eqnarray}
 \label{elossem3}
   {\Delta E_{rad}^{(1)}\over E}\approx && -
   {{\alpha_s C_F \mu^2 L^2}\over 4\lambda_gE}
   \left[\ln{2E\over \mu^2L} -0.048\right]\, ,
 \\
 \label{elossab3}
   {\Delta E_{abs}^{(1)}\over E}\approx &&
   {{\pi\alpha_s C_F}\over 3} {{LT^2}\over {\lambda_g E^2}}
   \left[
   \ln{{\mu^2L}\over T} -1+\gamma_{\rm E}-{{6\zeta^\prime(2)}\over\pi^2}
\right] .
 \end{eqnarray}
Our analytic approximation of the GLV zero-temperature result \cite{GLV}
also agrees with the improved limit by Zakharov \cite{Zakharov2}.
However, our result is accurate through the order of $1/E$. 
In Eq.~(\ref{elossab3}), we have assumed $\mu^2L/T\gg 1$ and kept only
the first two leading terms. In this limit, the average formation time 
for stimulated emission or thermal absorption is much smaller than the 
total propagation length. Therefore, the energy gain, 
$\Delta E_{abs}^{(1)}$, by thermal absorption (with partial 
cancellation by the stimulated emission) is linear in $L$, 
as compared to the quadratic dependence in the zero-temperature case.
However, the logarithmic dependence on $\mu^2L/T$ as compared to the
factor $\ln(4ET/\mu^2)$ in Eq.~(\ref{elossab0}) for no rescattering
is still a consequence of the LPM interference in medium. A quadratic 
$L$-dependence of $\Delta E_{abs}^{(1)}$ will arise
when $\mu^2L/T\ll 1$ \cite{WWlong}.

{\it Numerical results.} ---
To study the significance of the thermal absorption relative
to the induced radiation, we evaluate Eqs.~(\ref{eloss0}),
(\ref{elossem1}) and (\ref{elossab1}) numerically. We assume
the Debye screening mass to be $\mu^2=4\pi\alpha_s T^2$ from
the perturbative QCD at finite temperature \cite{HTL}. The
mean-free-path for a gluon $\lambda_g$ in the GW model is \cite{GW94},
 \begin{equation}
 \label{lambda}
    \lambda_g^{-1}= \langle\sigma_{qg}\rho_q\rangle
+\langle\sigma_{gg}\rho_g\rangle
    \approx {2\pi\alpha^2_s\over \mu^2} 9\times 7\zeta(3)
    {T^3\over \pi^2}\, ,
 \end{equation}
where $\zeta(3)\approx 1.202$. With fixed values of $L/\lambda_g$
and $\alpha_s$, $\Delta E/\mu$ should be a function of $E/\mu$ only.
\begin{figure}
\centerline{\psfig{figure=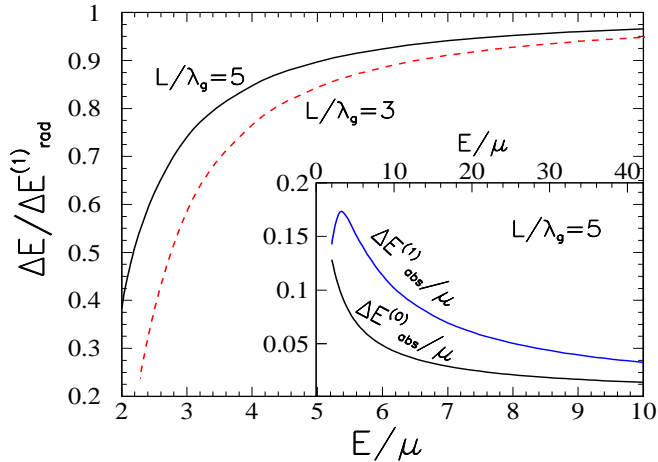,width=3.4in,height=2.4in}}
\vspace{0.15in}
\caption{The ratio of effective parton energy loss with
($\Delta E=\Delta E^{(0)}_{abs}+\Delta E^{(1)}_{abs}
+\Delta E^{(1)}_{rad}$) and without ($\Delta E^{(1)}_{rad}$)
absorption as a function of $E/\mu$. 
Inserted box: energy gain via absorption with 
($\Delta E^{(1)}_{abs}$) and without ($\Delta E^{(0)}_{abs}$)
rescattering.}
\label{fig1}
\end{figure}
Shown in Fig.~\ref{fig1} are ratios of the 
calculated radiative energy loss with and without stimulated
emission and thermal absorption as functions of $E/\mu$ for 
$L/\lambda_g=3$,5 and $\alpha_s=0.3$.
The thermal absorption reduces the effective parton energy loss 
by about 30-10\% for intermediate values of parton energy. This will 
increase the energy dependence of the effective parton energy loss
in the intermediate energy region. However, for partons with very 
high energy the effect of the gluon absorption is small and can be
neglected. Shown in the inserted box are the energy gain via
gluon absorption with ($\Delta E^{(1)}_{abs}$) and without
($\Delta E^{(0)}_{abs}$) rescattering.

{\it Conclusions.} --- In summary, we have considered the effect of
stimulated emission and thermal absorption in the calculation of the
effective energy loss for an energetic parton propagating in a
quark-gluon plasma. Even with partial cancellation by stimulated 
emission, the net result is an energy gain via absorption which
reduces the effective parton energy loss. Such a reduction is found
to be important for intermediate parton energy but can be neglected
at very high energies. For large distance $\mu^2L/T\gg 1$, we find
that the energy gain due to induced absorption with rescattering 
is linear in $L$, {\it modulo} a logarithmic dependence.
As in the case of QCD evolution of the
fragmentation functions in the vacuum, one should resum higher
order contributions of gluon
absorption and stimulated emission. Such a study in QED \cite{weldon}
has found the resummation to be important in the final radiation
spectrum. We expect the same in the QCD case with and
without multiple rescattering. This might further reduce the effective
parton energy loss in a hot QCD medium. 

{\it Acknowledgements.} ---
This work was supported by the National Natural Science Foundation
of China under projects 19928511 and 19945001, and by 
the Director, Office of Energy 
Research, Office of High Energy and Nuclear Physics, 
Division of Nuclear Physics, and by the Office of Basic Energy Science, 
Division of Nuclear Science, of  the U.S. Department of Energy 
under Contract No. DE-AC03-76SF00098. 
E.~W. thanks LBNL Nuclear Theory Group for their hospitality
during the completion of this work.


\end{multicols}
\end{document}